\documentstyle[11pt,aasms4,epsfig]{article}
\def\w{{\cal W}}
\def\ti{{ {\tilde \Delta}}}
\newcommand{\be}{\begin{equation}} 
\newcommand{\ee}{\end{equation}} 
\newcommand{\bea}{\begin{eqnarray}} 
\newcommand{\eea}{\end{eqnarray}}

\slugcomment{CERN/IC/TAC}
\lefthead{Contaldi, Magueijo,Ferreira, \& G\'orski}
\righthead{Cosmic Microwave Background}

\begin{document}
\tighten
\title{A Bayesian estimate of the skewness of the Cosmic Microwave Background}
\author{
C. R. Contaldi\altaffilmark{1},
P. G. Ferreira\altaffilmark{2,3,4},
J. Magueijo \altaffilmark{1},
and 
K. M. G\'orski\altaffilmark{5,6}}
\altaffiltext{1}{Theoretical Physics, 
Imperial College, Prince Consort Road, London SW7 2BZ, UK}
\altaffiltext{2}{Theory Group, CERN, CH-1211, Gen\`eve 23 ,Switzerland}
\altaffiltext{3}{CENTRA, Instituto Superior Tecnico, Lisbon, Portugal}
\altaffiltext{4}{Department de Physique Th\'eorique, Universit\'e de
Gen\`eve, CH-1211 Gen\`eve 4, Switzerland}
\altaffiltext{5}
{ESO, Karl Schwarzschild Strasse 2,
85748 Garching bei Mueanchen, Germany}
\altaffiltext{6}{
Warsaw University Observatory, Warsaw, Poland}
\begin{abstract}
We propose a formalism for estimating the skewness and angular
power spectrum of a general Cosmic Microwave Background data set.
We use the Edgeworth Expansion to define a
non-Gaussian likelihood function that
 takes into account the anisotropic nature of the
noise and the incompleteness of the sky coverage.
The formalism is then applied to estimate  the skewness of the 
publicly available 4 year Cosmic  Background
Explorer (COBE) Differential Microwave Radiometer data.  We find that
the data is consistent with a Gaussian skewness, and with isotropy.
Inclusion of non Gaussian degrees of freedom has essentially
no effect on
estimates of the power spectrum, if each $C_\ell$ is regarded 
as a separate parameter or if the angular power spectrum
is parametrized in terms of an amplitude ($Q$) and spectral index
($n$). 
Fixing the value of the angular power spectrum at its maxiumum
likelihood estimate, the  best fit skewness is 
$S=6.5\pm6.0\times10^4$($\mu$K)$^3$; marginalizing over $Q$ the estimate
of the skewness is  $S=6.5\pm8.4\times10^4$($\mu$K)$^3$ and
marginalizing over $n$ one has $S=6.5\pm8.5\times10^4$($\mu$K)$^3$.

\end{abstract}


\tighten
\section{Introduction}
\label{introduction}
The main assumption in current analysis of Cosmic Microwave
Background (CMB) data is that the perturbations responsible for
structure formation in the universe are Gaussian. The theoretical prejudice
is that the origin of these fluctuations was a period of
Inflation: the vacuum fluctuations of the inflaton field
on small scales (which satisfy Gaussian statistics) were linearly
 amplified  and stretched to super horizon scales through the 
superluminal expansion of the universe. Although there are
proposals for non-Gaussian fluctuations within the
framework of inflation
(\cite{srednicki,gangui,peebles,linde}),
 the
paradigm is suficiently compelling that the
assumption of Gaussianity has been unquestioned when
constructing methods of data analysis.

Recently, a number of groups have detected 
non-Gaussianity in the publicly available
4 year COBE DMR data. \cite{fmg}
 found that the COBE normalized bispectrum
was inconsistent with that of a Gaussian random field and
were unable to atribute it to systematic effects or foreground
contaminants.
\cite{pvl98}, using a statistic
defined in terms of the wavelet transform of the COBE data,
found evidence for non-Gaussianity localized in the
northern hemisphere and Novikov, Feldman and Shandarin (1998)
identified deviations from Gaussianity using topological
measures; \cite{bt99} have confirmed these results.
\cite{bzg99} have suggested that the culprit for this non-Gaussianity,
is a systematic
effect in the time ordered data used to construct the
publicly available COBE maps.

Possibly the most important result to come out of the
analysis of the 4 year COBE data is an estimate of 
the angular power spectrum of the CMB; it can be used
to normalize theories of large scale structure which
when then compared to measures of clustering on other
scales can be assessed as to their ability to 
describe our universe. In {\it all} estimates of the
angular spectrum, the fluctuations in the CMB are assumed
to be Gaussian. Given that the publicly available
 COBE 4 year data is indeed
non-Gaussian, there is a possibility that all
estimates of the angular power spectrum are biased in
some way. In this {\it letter} we present a new estimate
of the angular power spectrum using an approximate
likelihood function which includes non-Gaussian degrees of freedom.
In  section \ref{formalism} we introduce the Edgeworth expansion;
in section \ref{results} we apply it to the 4 year COBE DMR data
and in section \ref{discussion} we discuss our findings.

\section{The Formalism}
\label{formalism}
In most of what follows we will use the
notation of \cite{gorski94}. A map of the CMB as rendered by
the COBE DMR instrument, $\Delta_p$,  can be expanded in
spherical harmonics
\begin{equation}
\Delta_p=\sum_{\ell=0}^{\ell_{max}}\sum_{m=-\ell}^{\ell}
[(a^{CMB}_{\ell m}+a_{\ell m}^{G})w^{DMR}_{\ell}+a_{\ell m}^{N}]
w_{\ell}^{pix}Y_{\ell m}({\bf n}_p)
\end{equation} 
where $Y_{\ell m}$ are real, orthonormal spherical harmonics, 
$w_{\ell}^{DMR}$ are the filter coefficients of the DMR beam
pattern (Wright et al. 1994), and $w_\ell^{pix}$ are the
filter coefficients used to model smoothing due to pixelization;
$a^{CMB}_{\ell m}$, $a^{Gal}_{\ell m}$ and $a^{N}_{\ell m}$ are
the harmonic coefficients of the CMB anisotropy, Galactic emission
and instrument noise (from now one we will use the single
index notation for the spherical harmonic labels, i.e.
$i=\ell^2+\ell+1+m$); ${\bf n}_p$ is the unit vector which points
at pixel $p$. A major aim
of the DMR analysis is to obtain a good estimate of $a_{\ell m}^{MB}$,
separating out the effects due to the Galactic contamination and
reciever noise. This can be done by modeling the Galactic emission,
using information at different frequencies \cite{benn92},
and identifing which pixels in $\Delta_p$ are contaminated; one then
excises these pixels from the map.

Introducing a Galactic cut in the data renders the $Y_{\ell m}$s
non-orthonormal. \cite{gorski94} proposed a solution to this
problem: he identified a new basis, $\psi_{i}({\bf n}_p)$ which is orthonormal
on the cut sphere. The $a_{i}$ and the $Y_{i}({\bf n}_p$ can be
related to the harmonic coeffecients, $c_i$ and functions in the new
basis through ${\bf c}={\bf L}^T{\bf a}$ and
${\bf \psi}({\bf n}_p)={\bf L}^{-1}{\bf Y}({\bf n}_p)$.
We have introduced vector notation for quantities with index $i$;
${\bf L}$ is an triangular matrix so that $c_i$ is a combination
of $a_j$ with $j\ge i$. It is straightforward to see that
$\Delta_p={\bf c}^T\psi({\bf n}_p)$. 

If the sky is assumed to be Gaussian, then the probability
distribution of Fourier amplitudes is 
\begin{eqnarray}
G({\bf c})=\frac{d{\bf c}}{(2\pi)^{N_c/2}}\frac{e^{-(1/2){\bf c}^T
({\cal C}_{CMB}+{\cal C}_{N})^{-1}{\bf c}}}{\sqrt{\det
({\cal C}_{CMB}+{\cal C}_{N})}}
\end{eqnarray}
Here ${\cal C}_{CMB}=\langle{\bf c}_{CMB}{\bf c}_{CMB}^T\rangle=
{\bf L}^T\langle{\bf a}_{CMB}{\bf a}_{CMB}^T\rangle{\bf L}$,
$\langle{\bf a}_{CMB}{\bf a}_{CMB}^T\rangle={\rm diag}(a^2_i)$ and the 
noise covariance matrix is
${\cal C}_N=\Omega_{pix}\sum_{p\in \{cut sky\}}\sigma^2_p\psi({\bf n}_p)
\psi^T({\bf n}_p)$ (where $\sigma_p$ is
the RMS noise in pixel $p$ and $\Omega_{pix}$ is the solid angle of
the pixel). The Bayesian approach is, given the ${\bf c}$ to find
the $C_{\ell}=\langle a^2_{i(\ell,m)}\rangle$ which minimize the Likelihood function,
${\cal L}(C_\ell|{\bf C})=G({\bf c}|C_\ell)$ (where we have assumed
uniform priors for both ${\bf c}$ and $C_\ell$).

The Edgeworth expansion permits a systematic extension of
the  Likelihood function to include non-Gaussian degrees of
freedom (\cite{kso,chambers67,mccullagh84,roman95,amendola}); 
in this letter we shall assume
that the non-Gaussianity manifests itself as different levels of skewness,
${\cal S}_\alpha$, in sets of pixels, ${\cal P}_\alpha$. 
The Edgeworth expansion
for the skewness is
\begin{equation}
F(C_\ell,S_\alpha|{\bf c})=
G({\bf c}|C_\ell)[1+\sum_{\alpha}\frac{1}{6}S_\alpha\!\!\!
\sum_{p\in{\cal P}_\alpha
}\!\!\!{\cal E}^3_p+\sum_{\alpha}\sum_{\beta}
\frac{1}{72}S_\beta S_\alpha\!\!\!
\sum_{p'\in{\cal P}_\beta}\sum_{p\in{\cal P}_\alpha}\!\!\!{\cal
E}^6_{pp'}]
\label{edge1}
\end{equation}
where defining ${\w}=\psi({\bf n}_p)({\cal C}_{CMB}+{\cal
C}_{N})^{-1}\psi^T({\bf n}_{p'})$ and 
$\ti_p=\w_{pp'}{\Delta}_{p'}$ one has:
\begin{eqnarray}
{\cal E}^2_p&=&\ti_p-\w_{pp}\nonumber\\
{\cal E}^3_p&=&
\ti_p^3-3\w_{pp}\ti_p \nonumber \\
{\cal E}^6_{pp'}&=&
{\cal E}^3_p{\cal E}^3_{p'}-9{\cal E}^2_p\w_{pp'}{\cal E}^2_{p'}
+18\ti_p\w^2_{pp'}\ti_{p'}-6\w^2_{pp'}
\label{edge2}
\end{eqnarray}

The Edgeworth expansion contains a number of undesirable
properties. Seen as a function of the data, for fixed parameters,
a series truncation is in general not a density (not positive definite) and 
nor unimodal (not a single maximum). These properties follow
from the general form of the hermite polynomials. For fixed
data, as a function of the parameters, any truncation
still suffers from the same, and other equally serious, problems.
We work with a prescription which allows us to bypass this inconvenience.
We rewrite the likelihood as 
\bea
F({\bf x})=G({\bf x})\left[
\prod_{\alpha}
\prod_{p\in{\cal P}_\alpha}
(1+{S_\alpha\over 6  }{\cal E}^3_p
+{S_\alpha^2\over 72 }{\cal E}^6_p)\right]\times
\left[1+\sum_{\alpha}\sum_{\beta}{S_\alpha S_\beta\over 72 }(
\sum_{p\in{\cal P}_\alpha \neq  p'\in{\cal P}_\beta}
{\cal E}^6_{pp'}- {\cal E}^3_p {\cal E}^3_{p'})\right]
\eea
which is equivalent to (\ref{edge1}) up to second order.  The last
factor vanishes should there not be any correlations between pixels. 
We then expand the logarithm of the likelihood in a power series: 
\bea\label{exp2}
\log F= \log G 
&+&\sum_{\alpha}{S_\alpha\over 6  }\sum_{p\in{\cal P}_\alpha} {\cal E}^3_p
+\sum_{\alpha}{S^2_\alpha\over 72 }\sum_{p\in{\cal P}_\alpha}
\left[{\cal E}^6_p-({\cal E}^3_p)^2\right]\nonumber \\
&+&\sum_{\alpha}\sum_{\beta}
{S_\alpha S_\beta\over 72 }\sum_{p\in{\cal P}_\alpha \neq p'\in{\cal P}_\beta}
({\cal E}^6_{pp'}- {\cal E}^3_p {\cal E}^3_{p'})
\eea
One can check that exponentiation of (\ref{exp2}) differs from 
(\ref{edge1}) in terms higher than second order. 
However exponentiation of (\ref{exp2}) leads to a density. In fact
the likelihood as a function of $s$ is now a Gaussian distribution.
Our expansion is formally the same as the Edgeworth expansion;
however any truncation of a series of the form (\ref{exp2}) is very different
from a similar truncation of the Edgeworth series, and is better
behaved. This trick is described in [6.19] of \cite{kso}.
One may check that if the data is nearly Gaussian (in the sense that
the sample cumulants are small), and if there is
no noise, then our estimator is the trivial estimator
$\sum_p (\Delta^3_p)/N$. On the contrary (\ref{edge1}) fails to display
a maximum in most circumstances. 

The goal is then to minimize $-\log F$, given by Equations (\ref{exp2})
and (\ref{edge2}), in terms of $C_\ell$ and $S_\alpha$. A few comments
are in order. Firstly we have derived the most general Edgeworth
expansion in terms of skewness; a priori it would be natural 
to atribute one value of the skewness to all the pixels
in the cut sky (so there is only one $\alpha$ and ${\cal P}_\alpha$
is the set of all pixels in the map). However, a more conservative
attitude is to relax statistical isotropy and assume that different
parts of the map will have different levels of skewness.
In Section (\ref{results}) we will estimate the skewness of the
full sky and the skewness of the northern and southern hemispheres.
Secondly, the Edgeworth expansion is valid in the regime
of weak non-Gaussianity. Given that there is evidence of strong 
non-Gaussianity in the COBE one might expect that we are beyond the
regime of applicability of the formalism. However one must bear in
mind that the non-Gaussianity which has been identified has been
``localized'' in harmonic space; statistical tests in pixel
space have been consistent with the hypothesis the 4 year COBE DMR
data is a realization of a Gaussian random field.

\section{Results}
\label{results}

In this section we apply the formalism we have developed to the 4 year COBE
DMR data. 
We will be using the  inverse noise variance weighted, coadded maps of 
the 53A, 53B, 90A and 90B {\it COBE}-DMR channels at resolution 6, pixelized in
the galactic frame. 
We use the  
extended galactic cut of \cite{banday97}, and 
\cite{benn96} to remove most of the emission from the plane of the Galaxy.
We project out the components of the map corresponding to $\psi_i$
with $i=1,4$ which effectively removes the monopole and dipole in
the process and keep 961 coefficients.

The first questions we are interested in addressing is if
the assumption of zero skewness leads to a bias in the estimate
of the angular power spectrum. We have found that the individual
estimates of the $C_\ell$ are negligably affected by introducing
skewness; the 
introducion of the Edgeworth expansion introduces a 
roughly constant renormalization of the likelihood near the
peak and consequently the $C_\ell$ estimates and its errorbars
are affect by only few percent. It is convenient to parametrize
the angular power spectrum in terms of a normalization, 
$Q={\sqrt{5\over 4\pi}}a_2$, and a spectral index, $n$ such that
\be
C_\ell=a_2^2{\Gamma (\ell + {n-1\over 2})\Gamma ({9-n\over 2})
\over \Gamma (\ell + {5-n\over 2})\Gamma ({3+n\over 2})}
\ee
In Figure 1 we plot the likelihood contours for $Q$ and $n$,
marginalizing over the skewness. Any modification to
the Gaussian estimate is sufficiently small that 
the Gaussian likelihood contours and our estimates are
indistinguishable.
This is a promising result: weakening the assumption of Gaussianity
does not change in any  way existing estimates
of the angular power spectrum. 

We now focus on the value and distribution of the skewness.
Fixing the various  $C_\ell$ at the maximum we find that likelihood
as a function of the the skewness takes the form given in Figure 2
(solid line). 
By construction it is Gaussian with a peak and 1-$\sigma$ error
bars, $S=6.5\pm6.0\times10^4$($\mu$K)$^3$. The gaussian point ($S=0$)
lies well within the estimated probability distibution  i.e. our most
stringent estimate of skewness is consistent with
Gaussianity. 
One can relax the assumption of statistical
isotropy, and consider the skewness in the South and North 
hemispheres as separate parameters (dashed and dotted lines
respectively in Figure 2).  We observe that 
most of the statistical significance of our result comes
from pixels in the Southern hemisphere. There is no obvious reason
why this is so: the mean noise is essentially equivalent in
the northern and southern hemisphere as is its range of
values within each hemisphere.

An important point should be clarified with regards to
our estimate: the value we obtain is larger larger than what one
would expect by applying the naive estimator $S=\frac{1}{N_{pix}}
\sum_{p}\Delta_p^3$ as was used in Smoot {\it et al} 1994.
The latter estimate is only strictly valid if $\Delta_p$ are
uncorrelated, not the case of the level 6 pixelization of the COBE data.
In fact a rough estimate of the number of {\it effective}
(or uncorrelated) pixels indicates that the naive estimate should
be  a factor of 6-8 times smaller than what we get, 
which is indeed what we find.

One can attempt to identify correlations between the 
estimates of $Q$ or $n$ and $S$ by analysing the
joint likelihhod ($Q$,$S$) and ($n$,$S$) at the 
peak of the likelihood or marginalizing over $n$ and
$Q$ respectively. The results are presented in Figure
3 where we plot the $68\%$ and $95\%$ contour levels. 
Clearly there is very little correlation between
the pairs. Marginalizing over $n$ (Figure 3 a)
or $Q$ ( Figure 3 c) greatly increases the spread of
the likelihood in $Q$ and $n$ respecively but has essentially
the same effect on the effect on the estimated errors  of
the skewness; In the former case the maximum likelihood estimate
is $S=6.5\pm8.4\times10^4$($\mu$K)$^3$ and
marginalizing over $n$ one has $S=6.5\pm8.5\times10^4$($\mu$K)$^3$
Marginalizing over both $Q$ and
$n$ we obtain or most conservative estimate of the skewness,
$S= 6.5\pm8.7\times10^4$$\mu$K)$^3$.


\section{Discussion}
\label{discussion}
We set up a Baysian framework with which to study systematically 
non Gaussianity. This is the natural next step in power spectrum
estimation from CMB data sets with more general statistics
and should be incorporated in future algorithms. 
 We applied this framework to the
estimate of the skewness of publicly available COBE-DMR 4 year maps.
 We found that the
data is consistent with isotropy and Gaussianity as far as the skewness
is concerned. In agreement with this, the corrections on power
spectrum estimation due to introducing the skewness degree of freedom
are very small. If we parameterize the power spectrum in terms of
all the $C_\ell$ the effects are negligable and one can safely
use current {\it Gaussian} estimates of the angular power spectrum
as correct. We then focused on the various possible estimates
of the skewness and found our most conservative estimate to be 
$S=-6.5\pm 8.7 \times 10^4 \mu K^3$. 

We should perhaps comment on the differences between frequentist
and Bayesian studies of non Gaussianity. A frequentist approach
(such as the one used in Kogut {\it et al} (1996)) compares the
value of a statistic (say skewness) as measured in noisy data
with the distribution of what should have been measured if the
underlying signal were Gaussian. The value of the measured statistic
is {\it not} an estimate for its value for our sky. The Bayesian estimate
on the contrary, is an estimate of what the signal skewness actually
is, given the noise properties of the instrument, and what was observed.
The Bayesian estimate for the skewness and its frequentist value
need not, in fact, should not be the same. Nonetheless it is reassuring
that both answers agree in that the signal's skewness is Gaussian.

This is the first attempt at estimating the skewness from the
publicly available COBE-DMR 4 year maps; Kogut {\it et al} (1996)
estimated the pseudocollapsed and equilateral three point function
and found these to be also consistent with Gaussianity.
Our result reinforces the fact that the non-Gaussianity present
in COBE maps does not manifest itself in pixel space; the non-Gaussian
signal is mostly localized in frequency space and only statistics
that are sufficiently localized in $\ell$ space will be sensitive
to it. The natural next step is to extend the Edgeworth formalism
to the bispectrum; the complexity of the problem
makes it currently numerically intractable for general sky coverage and noise
covariance matrices. We are currently exploring various regimes
where such a formalism may be feasible.

\section*{Acknowledgments}

This work was performed on COSMOS, the Origin 2000 supercomputer owned 
by the UK-CCC and supported by HEFCE and PPARC; special thanks to
Stuart Rankin for endless patience. JM thanks the Royal Society 
for financial support. CRC is supported by the Beit Fellowship for
Scientific Research.

\begin{figure}
\centerline{\psfig{file=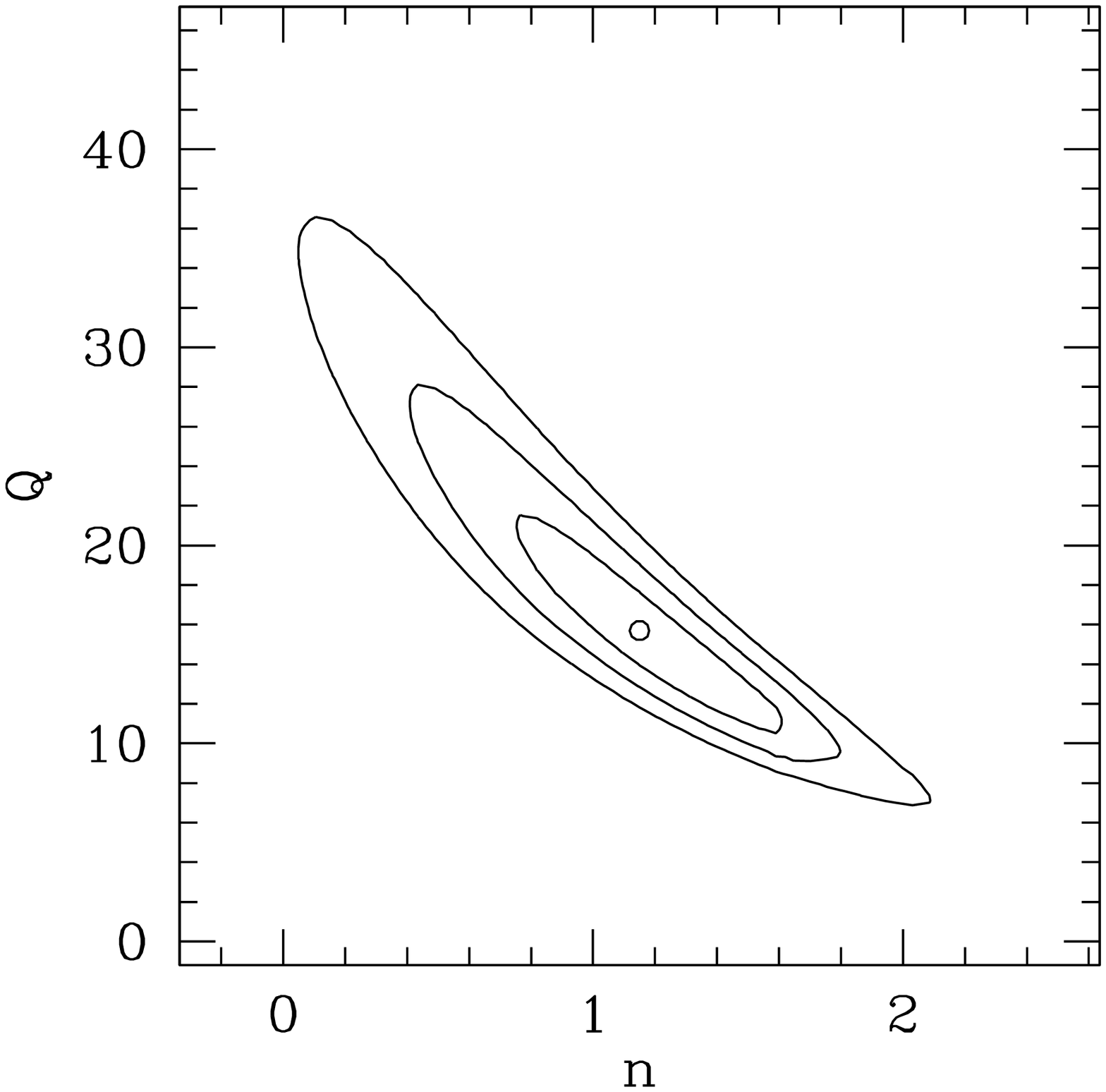,width=12 cm,angle=0}}
\caption{The likelihood as a function of the amplitude $Q$ and
spectral index $n$, marginalized over the skewness, $S$; the $68\%$,
$95\%$ and $99.7\%$ contours are plotted. The result is
indistinguishable from the $S=0$ likelihood.}
\label{Figure1}
\end{figure}

\begin{figure}
\centerline{\psfig{file=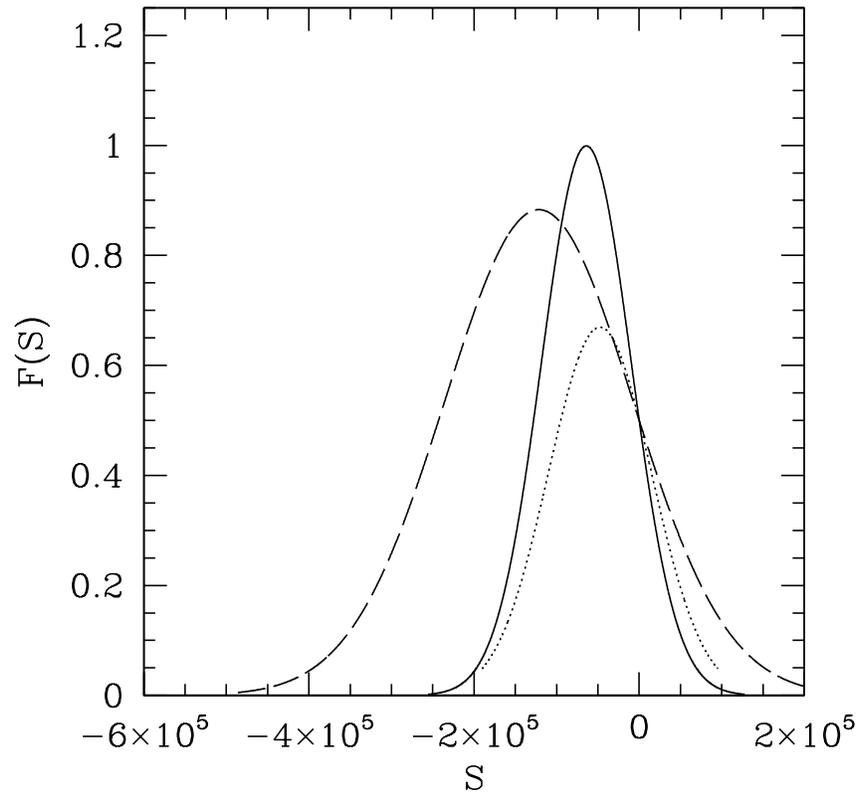,width=12 cm,angle=0}}
\caption{The Likelihood as a function of skewness, $S$ with
the angular power spectrum fixed at its maximum likelihood estimates;
the dashed line (dotted) line corresponds to using only the
northern (southern) hemisphere. The solid line corresponds to
using both.}
\label{Figure2}
\end{figure}

\begin{figure}
\centerline{\psfig{file=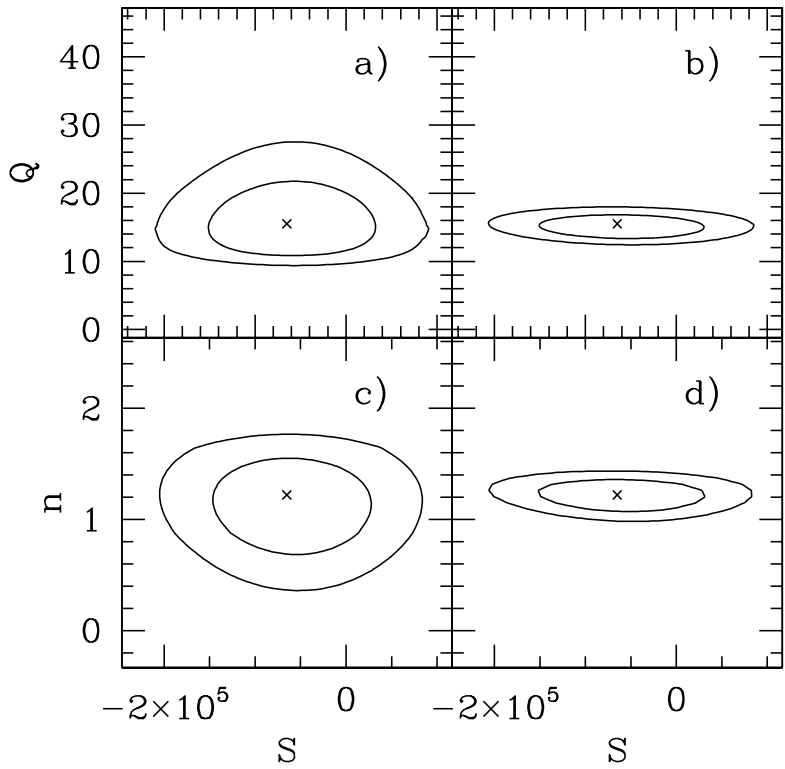,width=12 cm,angle=0}}
\caption{The likelihoods as a function of amplitude, $Q$ and skewness,
 $S$, a) marginalized over the spectral index, $n$ and b) with  $n$ fixed
at its maximum likelihood estimate.The likelihoods as a function of
 $n$ and $S$,
 c) marginalized over $Q$  and d)  with $Q$ fixed
at its maximum likelihood estimate.}
\label{Figure3}
\end{figure}


\begin{thebibliography}{99}
\bibitem[Amendola 1996]{amendola} Amendola, L. {\it M.N.R.A.S.} {\bf
283} 983 (1996).
\bibitem[Banday {\it et al} 1997]{banday97} Banday, A.J. {\it et al}
{\it Ap.J.}, {\bf 475}, 393 (1997).
\bibitem[Banday, Zaroubi \& G\'orski (1999)]{bzg99}Banday, A.J., Zaroubi, S.,
G\'orski, K.M. {\tt astro-ph/9908070}
\bibitem[Bennet {\it et al} (1992)]{benn92} Bennet, C.L. {\it et al}
{\it Ap.J.} {\bf 396} L7 (1992).
\bibitem[Bennet {\it et al} 1996]{benn96} Bennet, C.L. {\it et al}
{\it Ap. J.} {\bf 464}, 1 (1996).
\bibitem[Bromley \& Tegmark]{bt99} Bromley, B. and Tegmark, M.
{\tt astro-ph/9904254}.
\bibitem[Chambers 1967]{chambers67} Chambers, J., {\it Biometrika}
{\bf 54} 367 (1967)
\bibitem[Ferreira, Magueijo \& G\'orski (1998)]{fmg}Ferreira, P.G.,
Magueijo, J. and G\'orski, K.M.G. {\it Ap.J.} {\bf 503} L1
(1998)
\bibitem[Gangui {\it et al} 1994]{gangui}Gangui, A.,
 Lucchin, F.,
 Mataresse, S. and
 Mollerach, S., {\it Ap.J} {\bf 430} 447 (1994).
\bibitem[G\'orski (1994)]{gorski94}G\'orski, K.M.G, {\it Ap.J} {\bf 430} L85 (1994) 
\bibitem[Juszkiewicz {\it et al} 1995]{roman95} Juszkiewicz, R. {\it et al} {\it Ap. J.}
{\bf 442} 39 (1995)
\bibitem[Kendall $\&$ Stuart 1977]{kso}Kendall, M.G. and Stuart, A.,
{\it The Advanced Theory of Statistics}, Charles Griffin (1977).
\bibitem[Kogut {\it et al} 1996]{kog96}Kogut, A. {\it et al}
{\it Ap.J.} {\bf 464} L29 (1996). 
\bibitem[Linde \& Mukhanov 1997]{linde}Linde, A. and Mukhanov, V.
{\it Phys.Rev.} {\bf D56} 535 (1997)
\bibitem[McCullagh 1984]{mccullagh84}McCullagh, p. {\it Biometrika}
{\bf 71} 461 (1984) 
\bibitem[Novikov, Feldman and Shandarin (1998)]{nfs98}
Novikov, D., Feldman, H. and Shandarin, S. {\tt astro-ph/9809238}
\bibitem[Pando, Valls-Gabaud \& Fang (1998)]{pvl98}Pando, J.,
Valls-Gabaud and Fang, L.Z., {\it Phys.Rev.Lett} {\bf 81} 4568 (1998).
\bibitem[Peebles 1997]{peebles}Peebles, P.J. 
{\it Ap.J.} {\bf 483}, L1 (1997).
\bibitem[Smoot {\it et al} 1994]{sm94} Smoot, G.  {\it et al} {\it Ap.
J.} {\bf 437} 1 (1994).
\bibitem[Srednicki 1993]{srednicki}Srednicki, M {\it Ap.J} {\bf 416}
L1 (1993).
\bibitem[Wright {\it et al} (1992)]{wright}Wright, E.L. {\it et al}
{\it Ap.J.} {\bf 396} L13 (1992).

\end{thebibliography}
\end{document}